\documentstyle[12pt]{article}                    
\newfont{\ffont}{msym10}                        
\newcommand{\beq}{\begin{equation}}             
\newcommand{\eeq}{\end{equation}}               
\newcommand{\bqry}{\begin{eqnarray}}            
\newcommand{\eqry}{\end{eqnarray}}              
\newcommand{\bqryn}{\begin{eqnarray*}}          
\newcommand{\eqryn}{\end{eqnarray*}}            
\newcommand{\NL}{\nonumber \\}                  
\newcommand{\preprint}[1]{\begin{table}[t]      
            \begin{flushright}                  
            \begin{large}{#1}\end{large}        
            \end{flushright}                    
            \end{table}}                        
\newcommand{\PD}[2]                             
    {\frac{\partial^{#2}}{\partial #1^{#2}}}    
\begin{document}
\preprint{LA-UR-96-2461 \\ IASSNS-HEP-96/82}
\title{On the Thermodynamics of \\ Chiral Symmetry Restoration}
\author{\\ L. Burakovsky\thanks{Bitnet: BURAKOV@QCD.LANL.GOV} \
\\  \\  Theoretical Division, T-8 \\  Los Alamos National  
Laboratory \\ Los
Alamos NM 87545, USA \\  \\  and  \\  \\
L.P. Horwitz\thanks{Bitnet: HORWITZ@SNS.IAS.EDU. On sabbatical leave from
School of Physics and Astronomy, Tel Aviv University, Ramat Aviv, Israel.
Also at Department of Physics, Bar-Ilan University, Ramat-Gan,  
Israel  } \
\\  \\ School of Natural Sciences \\ Institute for Advanced Study  
\\ Princeton
NJ 08540, USA \\}
\date{ }
\maketitle
\begin{abstract}
The formulas for the temperature dependences of the non-strange and  
strange
quark condensates are derived by taking into account the  
contribution of the
massive resonances. Critical temperature of the chiral symmetry  
restoration
transition is established to be 190 MeV if only meson resonances are 
considered, and 175 MeV if both meson and baryon resonances are  
taken into
account.
\end{abstract}

\bigskip
\section*{  }
For massless quarks, the QCD Hamiltonian is invariant under chiral
transformations. The phenomenology of the strong interactions is  
consistent
with this property provided that chiral symmetry is spontaneously  
broken.
Lattice gauge calculations \cite{lattice} indicate that, at low  
temperatures,
the scalar currents $\bar{q}q=\{\bar{u}u,\bar{d}d,\bar{s}s\}$  
develop nonzero
expectation values $\langle \bar{q}q\rangle _T$ which are referred  
to as quark
condensates, and represent order parameter of spontaneously broken chiral
symmetry. Chiral perturbation theory \cite{pert,GL,GL1} shows that  
$\langle
\bar{q}q\rangle _T$ melts as the temperature increases. One  
generally expects
that if the temperature reaches a critical value, $T_c,$ chiral  
symmetry is
restored. Lattice calculations \cite{KSW} confirm this expectation  
but show
that the strange quark condensate $\langle \bar{s}s\rangle $ does  
not melt in
the chiral restoration transition, supporting the description of  
this phase
transition in terms of the behavior of the $SU(2)\times SU(2)$ linear 
$\sigma $-model initiated by Wilczek {\it et al.} \cite{Wil}.  
Presumably,
deconfinement transition which liberates color, takes place at  
temperature of
the order of $T_c.$ At low enough temperatures, chiral symmetry provides 
constraints on the temperature dependences of several physical  
quantities. For
example, for the pion gas without interactions, one finds \cite{GL} 
\beq
\frac{\langle \bar{q}q\rangle _T}{\langle \bar{q}q\rangle  
_0}=1-\frac{N_f^2-1}{
3N_f}\frac{T^2}{4f_\pi ^2},\;\;\;
\frac{f_\pi (T)}{f_\pi }=1-\frac{N_f}{6}\frac{T^2}{4f_\pi ^2},\;\;\; 
\frac{\mu _\pi (T)}{m_\pi }=1+\frac{1}{6N_f}\frac{T^2}{4f_\pi ^2},
\eeq
where $f_\pi $ is the pion decay constant, $\simeq 93$ MeV, $\mu  
_\pi $ the
screening pion mass, and $N_f$ the number of massless flavors. In a  
real world,
$N_f=2,$ so that the first relation of (1) may be rewritten as
\beq
\frac{\langle \bar{q}q\rangle _T}{\langle \bar{q}q\rangle  
_0}=1-\frac{T^2}{8f_
\pi ^2}\simeq 1 - \left(\frac{T}{260\;{\rm MeV}}\right)^2.
\eeq
If one goes beyond the low temperature limit, predictions start to  
become
model dependent.

In this paper we calculate the correction to the formula for the  
non-strange
quark condensate, Eq. (2), given by the contribution of the higher mass 
hadrons, and obtain a similar relation for the strange quark  
condensate. It is
well known that the correct thermodynamic description of hot  
hadronic matter
requires consideration of higher mass excited states, the  
resonances, whose
contribution becomes essential at temperatures $\sim O(100$ MeV)
\cite{Shu,Leut,Bebie}. The method for taking into account these  
resonances was
suggested by Belenky and Landau \cite{BL} as considering unstable  
particles on
an equal footing with the stable ones in the thermodynamic  
quantities; e.g.,
the formulas for the pressure and energy density in a resonance gas 
read\footnote{Since the temperatures we are dealing with are much  
less than
the nucleon mass, we may treat all the particles as bosons, and  
distinguish
fermions only by the factor 7/8 in the expression for the particle  
degeneracy.
We also neglect chemical potential for simplicity.}
\beq
p=\sum _ip_i=\sum _ig_i\frac{m_i^2T^2}{2\pi ^2}\sum _{r=1}^\infty  
(\pm 1)^{r+1}
\frac{K_2(rm_i/T)}{r^2},
\eeq
\beq
\rho =\sum _i\rho _i,\;\;\;\rho _i=T\frac{dp_i}{dT}-p_i,
\eeq
where $+1$ $(-1)$ corresponds to the Bose-Einstein (Fermi-Dirac)  
statistics,
and $g_i$ are the corresponding degeneracies $(J$ and $I$ are spin and
isospin, respectively), $$g_i=\left[
\begin{array}{ll}
(2J_i+1)(2I_i+1) & {\rm for\;non-strange\;mesons} \\
4(2J_i+1) & {\rm for\;strange}\;(K)\;{\rm mesons} \\
2(2J_i+1)(2I_i+1) & {\rm for\;baryons,}
\end{array} \right. $$ and since chiral symmetry suppresses the  
interactions of
low energy Goldstone bosons both among themselves and with massive  
hadrons, the
gas can be approximately described as a collection of free particles. 

To derive an explicit relation for the value of the quark condensate as a
function of temperature, we first note that the operator $\bar{q}q$  
occurs in
the quark mass term of the Hamiltonian,
\beq
H=H_0+\int d^3x\sum _{q=u,d,s}m_q\;\bar{q}q,
\eeq
where $H_0$ is the Hamiltonian of massless (chirally symmetric) QCD. The 
thermal expectation value of $\bar{q}q$ represents, therefore, the  
response of
the partition function to a change in the quark mass,
\beq
\langle \bar{q}q\rangle _T=-\frac{1}{V}\frac{\partial \ln  
Z}{\partial m_q},
\eeq
where $V$ is the four-dimensional euclidean volume $(=V^{(3)}/T).$ At
sufficiently large volume, $\ln Z\rightarrow V(p-\rho _0),$ where  
$p$ is the
pressure and $\rho _0$ the vacuum energy density. In the large  
volume limit,
Eq. (6) therefore takes on the form
\beq
\langle \bar{q}q\rangle _T=\frac{\partial \rho _0}{\partial m_q}-
\frac{\partial p}{\partial m_q}=\langle \bar{q}q\rangle _0-
\frac{\partial p}{\partial m_q}.
\eeq
Since $p$ depends on $m_q$ only through the masses of the particles, one
obtains from (3),(7), through $d/dx\;(x^2K_2(x))=-x^2K_1(x),$
\beq
\langle \bar{q}q\rangle _T=\langle \bar{q}q\rangle _0+\sum  
_ig_i\frac{m_i^2T}{2
\pi ^2}\frac{\partial m_i}{\partial m_q}\sum _r(\pm 1)^{r+1}
\frac{K_1(rm_i/T)}{r}.
\eeq

1) Non-strange quark condensate. \\ One uses the lowest order relation 
\cite{GOR}
\beq
m_\pi ^2f_\pi ^2=-m\langle \bar{q}q\rangle _0,
\eeq
where $\langle \bar{q}q\rangle _0\equiv \langle  
\bar{u}u+\bar{d}d\rangle _0$
and $m\equiv 1/2\;(m_u+m_d),$ and obtains from (8),
\beq
\langle \bar{q}q\rangle _T=\langle \bar{q}q\rangle  
_0\left(1-\frac{T^2}{f_\pi
^2}\sum _i\frac{g_i}{2\pi ^2}\frac{m_im}{m_\pi ^2}\frac{\partial m_i}{
\partial m}\sum _r(\pm  
1)^{r+1}\frac{m_i}{rT}K_1\left(\frac{rm_i}{T}\right)
\right).
\eeq

1) Strange quark condensate. \\ Now one uses the lowest order  
relation for the
$K$ meson $(f_K$ being the kaon decay constant, $\simeq 114$ MeV),  
\beq
m_K ^2f_K ^2=-1/2\;(m+m_s)\langle \bar{s}s+1/2\;\bar{q}q\rangle  
_0\cong -m_s
\langle \bar{s}s\rangle _0,
\eeq
since $m_s>>m,$ and $\langle \bar{s}s\rangle _0\simeq \langle  
\bar{u}u\rangle
_0\simeq \langle \bar{d}d\rangle _0$ \cite{SVZ}, and obtains from (8),
\beq
\langle \bar{s}s\rangle _T=\langle \bar{s}s\rangle  
_0\left(1-\frac{T^2}{f_K
^2}\sum _i\frac{g_i}{2\pi ^2}\frac{m_im_s}{m_K ^2}\frac{\partial m_i}{
\partial m_s}\sum _r(\pm  
1)^{r+1}\frac{m_i}{rT}K_1\left(\frac{rm_i}{T}\right)
\right).
\eeq
Gerber and Leutwyler \cite{GL1} have established the $T$-dependence  
of the
non-strange quark condensate by direct calculation of the sum in Eq. (10)
for known resonances, using the estimate $\partial m_i/\partial  
m\simeq N_i,$
the number of valence quarks of type $u$ and $d,$ and $m\simeq 7$  
MeV. The
results of Gerber and Leutwyler predict chiral symmetry to be restored at
$T_c\simeq 200$ MeV if only massive states are taken into account,  
and with
the pion contribution in addition to that of the massive states,  
$T_c\simeq
190$ MeV for nonzero masses of the $u$- and $d$-quarks, and 170 MeV  
for zero
masses of the latter. In this paper we wish to obtain analytic  
expressions for
both the $T$-dependent non-strange and strange quark condensates,  
and compare
the results with those of Gerber and Leutwyler.

In the following we shall restrict ourselves to the meson  
resonances alone.
Baryon resonances may be treated in a way similar to that described  
in this
paper, with the introduction of two chemical potentials, for both  
conserved
net baryon number and strangeness. We shall dwell briefly on this  
point at
the end of the paper.

To calculate the derivatives $\partial m_i/\partial m_q$ in Eqs.  
(10),(12), we
first note that the expansion of $m_\pi ^2$ and $m_K^2$ in the  
powers of $m_u,
m_d,m_s$ starts with terms which are linear in the quark masses,
\beq
m_\pi ^2=2mB,\;\;\;m_K^2=(m+m_s)B.
\eeq
For higher mass meson resonances, we expect that similar relations  
hold (with
$C$ being constant within a given meson nonet):
\beq
m_{0^{''}}^2\simeq  
m_1^2=2mB+C,\;\;\;m_{1/2}^2=(m+m_s)B+C,\;\;\;m_{0^{'}}^2
\simeq 2m_sB+C,
\eeq
where $m_1,m_{1/2},m_{0^{'}},m_{0^{''}}$ are the masses of the isovector,
isospinor and two isoscalar states, respectively, and $0^{'}$  
belongs to a
mostly octet. Indeed, for the vector meson nonet these relations  
may be easily
obtained from the constancy of the differences of the squared  
masses of the
corresponding spin-triplet and spin-singlet states \cite{Lucha},  
$$\triangle
M^2\equiv M^2(^3S_1)-M^2(^1S_0),$$ resulting in\footnote{Also,  
$m_{D^\ast }^2-
m_D^2=0.55$ GeV$^2,$ $m_{D^\ast _s}^2-m_{D_s}^2=0.58$ GeV$^2,$  
$m_{B^\ast }^2-
m_B^2=0.55$ GeV$^2,$ with the exception of the $c\bar{c}$ states  
for which
$m_{J/\psi }^2-m_{\eta _c}^2=0.70$ GeV$^2.$}
\beq
m_\rho ^2-m_\pi ^2=0.57\;{\rm GeV}^2,\;\;\;m_{K^\ast }^2-m_K^2=0.55\;
{\rm GeV}^2,
\eeq
and two Gell-Mann--Okubo mass formulas, the standard one \cite{GMO},
\beq
m_1^2+3m_8^2=4m_{1/2}^2,
\eeq
and an extra relation \cite{su4,linear}
\beq
m_{0^{'}}^2+m_{0^{''}}^2=m_8^2+m_0^2=2m_{1/2}^2,
\eeq
with $m_0$ and $m_8$ being the masses of the isoscalar octet and singlet 
states, respectively, which for an almost ideally mixed nonet reduces to 
\cite{su4,linear}
\beq
m_{0^{''}}^2\simeq m_1^2,\;\;\;m_{0^{'}}^2\simeq 2m_{1/2}^2-m_1^2.
\eeq
Therefore, it follows from (13),(15),(18) that $$m_\omega ^2\simeq  
m_\rho ^2=
2mB+C,\;\;\;m_{K^\ast }^2=(m+m_s)B+C,\;\;\;m_\phi ^2\simeq  
2m_sB+C,$$ with
$C\simeq 0.56$ GeV$^2.$ It was shown by Bal\'{a}zs \cite{Bal} that  
$m_\rho ^2-
m_\pi ^2=1/2\alpha ^{'},$ with $\alpha ^{'}$ being a universal  
Regge slope,
$\alpha ^{'}\simeq 0.85$ GeV$^{-2}$, in agreement with (15), so that
\beq
m_\omega ^2\simeq m_\rho ^2=2mB+1/2\alpha ^{'},\;\;\;m_{K^\ast  
}^2=(m+m_s)B+
1/2\alpha ^{'},\;\;\;m_\phi ^2\simeq 2m_sB+1/2\alpha ^{'}.
\eeq
For higher spin nonets, since the corresponding states with equal  
isospin and
alternating parity lie on the linear Regge trajectories, one has, e.g.,
\bqry
m_{a_2}^2 & = & m_\rho ^2+1/\alpha ^{'}\;=\;2mB+3/2\alpha ^{'}, \NL  
m_{K_2^\ast }^2 & = & m_{K^\ast }^2\!+1/\alpha  
^{'}\!=\;(m+m_s)B+3/2\alpha ^{
'}, \NL
m_{f_2}^2 & = & m_\omega ^2+1/\alpha ^{'}\;\simeq \;2mB+3/2\alpha  
^{'}, \NL
m_{f_2^{'}}^2 & = & m_\phi ^2+1/\alpha ^{'}\;\simeq  
\;2m_sB+3/2\alpha ^{'},
\eqry
\bqry
m_{\rho _3}^2 & = & m_{a_2}^2+1/\alpha ^{'}\;=\;2mB+5/2\alpha ^{'}, \NL  
m_{K_3^\ast }^2 & = & m_{K_2^\ast }^2+1/\alpha  
^{'}=\;(m+m_s)B+5/2\alpha ^{'},
\NL
m_{\omega _3}^2 & = & m_{f_2}^2+1/\alpha ^{'}\;\simeq  
\;2mB+5/2\alpha ^{'},
\NL
m_{\phi _3}^2 & = & m_{f_2^{'}}^2+1/\alpha ^{'}\;\simeq  
\;2m_sB+5/2\alpha ^{'},
\;\;\;{\rm etc.,}
\eqry
and also
\bqry
m_{b_1}^2 & = & m_\pi ^2+1/\alpha ^{'}\;=\;2mB+1/\alpha ^{'}, \NL
m_{K_1}^2 & = & m_K^2+1/\alpha ^{'}\;\!=\;(m+m_s)B+1/\alpha  
^{'},\;\;\;{\rm
etc.}
\eqry
Thus, we consider the relations (14) as granted by both the  
Gell-Mann--Okubo
mass formula for a close-to-ideally mixed nonet and the Regge  
phenomenology.
It then follows from (13),(14) that
\bqry
m\frac{\partial m_{0^{''}}}{\partial m} & \simeq & m\frac{\partial m_1}{
\partial m}\;=\;\frac{mB}{m_1}\;=\;\frac{m_\pi ^2}{2m_1}, \NL
m\frac{\partial m_{1/2}}{\partial m} & = & \frac{mB}{2m_{1/2}}\;=\;\frac{
m_\pi ^2}{4m_{1/2}}, \NL
m\frac{\partial m_{0^{'}}}{\partial m} & \simeq & 0,
\eqry
and
\bqry
m_s\frac{\partial m_{0^{''}}}{\partial m_s} & \simeq &  
m_s\frac{\partial m_1}{
\partial m_s}\;=\;0, \NL
m_s\frac{\partial m_{1/2}}{\partial m_s} & = & \frac{m_sB}{2m_{1/2}}\;=\;
\frac{m_sm_K^2}{2m_{1/2}(m+m_s)}\;\cong \;\frac{m_K^2}{2m_{1/2}}, \NL 
m_s\frac{\partial m_{0^{'}}}{\partial m_s} & \simeq &  
\frac{m_sB}{m_{0^{'}}}\;
=\;\frac{m_sm_K^2}{m_{0^{'}}(m+m_s)}\;\cong \;\frac{m_K^2}{m_{0^{'}}},
\eqry
since $m_s>>m.$ Thus, we find that the expressions to be inserted  
into Eqs.
(10) and (12) are, respectively,
\beq
m\frac{\partial m_i}{\partial m}=\frac{m_\pi ^2}{2m_i},\;\;\;
m_s\frac{\partial m_i}{\partial m_s}=\frac{m_K^2}{m_i},
\eeq
and out of 9 isospin degrees of freedom of a nonet, 6 contribute to  
the formula
for the non-strange condensate: 3 isovector, 2 isospinor and 1  
isoscalar which
belongs to a mostly singlet, and 3 to the formula for the strange  
condensate:
2 isospinor and 1 isoscalar which belongs a mostly  
octet.\footnote{In fact,
each of the $\bar{u}u,\bar{d}d,\bar{s}s$ condensates gains the  
contribution of
3 isospin degrees of freedom of a nonet.} We shall make another  
simplification
in Eqs. (10),(12), viz., approximate the Bose-Einstein statistics by the 
Maxwell-Boltzmann one, taking account of the sum over $r$ through  
the factor
$\pi ^2/6=\zeta (2)\equiv \sum _r1/r^2,$ which is the asymptotic  
form of this
sum for $T>>m_i.$ Then, the formulas (10) and (12) will finally reduce to
\beq
\langle \bar{q}q\rangle _T=\langle \bar{q}q\rangle  
_0\left(1-\frac{T^2}{24f_
\pi ^2}\sum _i ^{'}g_i\frac{m_i}{T}K_1\left(\frac{m_i}{T}\right)\right),
\eeq
\beq
\langle \bar{s}s\rangle _T=\langle \bar{s}s\rangle  
_0\left(1-\frac{T^2}{12f_K^
2}\sum _i ^{''}g_i\frac{m_i}{T}K_1\left(\frac{m_i}{T}\right)\right),
\eeq
where the primes indicate the summation over 6 and 3 isospin  
degrees of freedom
of a nonet, respectively. Now it is seen in Eq. (26) that if one  
restricts
himself to the massless pions alone, $g_\pi =3,$ one obtains the  
formula (2).
This formula sets the temperature scale for the non-strange  
condensate $\sim $
250 MeV. As seen in Eqs. (26),(27), the massive states accelerate  
the melting
of the condensates, so that the temperature scale for, at least, the 
non-strange condensate is expected to be much narrower.

To calculate the sums in Eqs. (26),(27), we resort to the notion of a  
{\it resonance spectrum} which is introduced in order to substitute the
summation over individual particle species by the integration over  
the mass in
the expressions for thermodynamic quantities, so that, e.g., Eqs.  
(3),(4) may
be rewritten (in the Maxwell-Boltzmann approximation for the particle 
statistics) as
\beq
p=\int _{m_1}^{m_2}dm\;\tau (m)p(m),\;\;\;p(m)\equiv  
\frac{m^2T^2}{2\pi ^2}
K_2\left(\frac{m}{T}\right),
\eeq
\beq
\rho =\int _{m_1}^{m_2}dm\;\tau (m)\rho (m),\;\;\;\rho (m)\equiv
T\frac{dp(m)}{dT}-p(m),
\eeq
and the resonance spectrum $\tau (m)$ is normalized as
\beq
\int _{m_1}^{m_2}dm\;\tau (m)=\sum _ig_i,
\eeq
where $m_1$ and $m_2$ are the masses of the lightest and heaviest  
species,
respectively, entering the formulas (3),(4).

In both the statistical bootstrap model \cite{Hag,Fra} and the dual  
resonance
model \cite{FV}, a resonance spectrum takes on the form
\beq
\tau (m)\sim m^a\;e^{m/T_0},
\eeq
where $a$ and $T_0$ are constants. The treatment of a hadronic  
resonance gas
by means of the spectrum (31) leads to a singularity in the  
thermodynamic
functions at $T=T_0$ \cite{Hag,Fra} and, in particular, to an  
infinite number
of the effective degrees of freedom in the hadron phase, thus hindering a
transition to the quark-gluon phase. Moreover, as shown by Fowler  
and Weiner
\cite{FW}, an exponential mass spectrum of the form (31) is  
incompatible with
the existence of the quark-gluon phase: in order that a phase  
transition from
the hadron phase to the quark-gluon phase be possible, the hadronic  
spectrum
cannot grow with $m$ faster than a power.

In our previous work \cite{spectrum} we considered a model for a  
transition
from a phase of strongly interacting hadron constituents, described by a 
manifestly covariant relativistic statistical mechanics which  
turned out to be
a reliable framework in the description of realistic physical systems 
\cite{mancov}, to the hadron phase described by a resonance  
spectrum, Eqs.
(28),(29). An example of such a transition may be a relativistic high 
temperature Bose-Einstein condensation studied by the authors in ref. 
\cite{cond}, which corresponds, in the way suggested by Haber and Weldon 
\cite{HW}, to spontaneous flavor symmetry breakdown, $SU(3)_F\rightarrow 
SU(2)_I\times U(1)_Y,$ upon which hadronic multiplets are formed,  
with the
masses obeying the Gell-Mann--Okubo formulas \cite{GMO}
\beq
m^\ell =a+bY+c\left[ \frac{Y^2}{4}-I(I+1)\right];
\eeq
here $I$ and $Y$ are the isospin and hypercharge, respectively,  
$\ell $ is 2
for mesons and 1 bor baryons, and $a,b,c$ are independent of $I$  
and $Y$ but,
in general, depend on $(p,q),$ where $(p,q)$ is any irreducible  
representation
of $SU(3).$ Then only the assumption on the overall degeneracy  
being conserved
during the transition is required to lead to the unique form of a  
resonance
spectrum in the hadron phase:
\beq
\tau (m)=Cm,\;\;\;C={\rm const}.
\eeq
Zhirov and Shuryak \cite{ZS} have found the same result on  
phenomenological
grounds. As shown in ref. \cite{ZS}, the spectrum (33), used in the  
formulas (28),(29) (with the upper limit of integration infinity), leads
to the equation of state $p,\rho \sim T^6,$ $p=\rho /5,$ called by  
Shuryak the
``realistic'' equation of state for hot hadronic matter \cite{Shu},  
which has
some experimental support. Zhirov and Shuryak \cite{ZS} have calculated 
the velocity of sound, $c_s^2\equiv dp/d\rho =c_s^2(T),$ with $p$  
and $\rho $
defined in Eqs. (3),(4), and found that $c_s^2(T)$ at first  
increases with $T$
very quickly and then saturates at the value of $c_s^2\simeq 1/3$  
if only the
pions are taken into account, and at $c_s^2\simeq 1/5$ if  
resonances up to
$M\sim 1.7$ GeV are included.

We have checked the coincidence of the results given by the linear  
spectrum
(33) with those obtained directly from Eq. (3) for the actual  
hadronic species
with the corresponding degeneracies, for all well-established hadronic 
multiplets, both mesonic and baryonic, and found it excellent  
\cite{spectrum}.
Therefore, the theoreticalconclusion that a linear spectrum is the  
actual
spectrum in the description of individual hadronic multiplets finds its 
experimental confirmation as well. In our recent papers  
\cite{su4,linear} we
have shown that a linear spectrum of an individual nonet is  
consistent with
the Gell-Mann--Okubo mass formula (16) (in fact, this formula may  
be derived
with the help of a linear spectrum \cite{su4}), and leads to an  
extra relation
for the masses of the isoscalar states, Eq. (17), which was checked  
in ref.
\cite{linear} and shown to hold with an accuracy of up to $\sim  
$3\% for all
well-established nonets. In ref. \cite{su4} we have generalized a linear 
spectrum to the case of four quark flavors and derived the corresponding 
Gell-Mann--Okubo mass formula for an $SU(4)$ meson hexadecuplet, in good
agreement with the experimentally established masses of the charmed  
mesons.
In ref. \cite{enigmas} we have applied a linear spectrum to the  
problem of
establishing the correct $q\bar{q}$ assignment for the problematic meson 
nonets, like the scalar, axial-vector and tensor ones, and  
separating out
non-$q\bar{q}$ mesons. In this paper we shall apply a resonance  
spectrum to
the derivation of the formulas for the temperature dependences of  
the quark
condensates.

It was shown in ref. \cite{spectrum} that the actual resonance  
spectrum does
not depend on the dimensionality of spacetime.\footnote{This is another 
argument against the Hagedorn spectrum, since the exponent $a$ in  
Eq. (31)
depends explicitly on the dimensionality of spacetime (it is  
related to the
number of transverse dimensions of a string theory \cite{DC}).}  
Therefore, the
sums in Eqs. (26),(27) which are, in fact, related to the  
expression for the
pressure of a free gas in 1+1 dimensions \cite{1+1}, as calculated for 
individual nonets, may be substituted by the integration over $m$ with a 
linear spectrum. The normalization constant $C$ was established in ref. 
\cite{spectrum}: for a nonet, one has 9 isospin degrees of freedom  
lying in
the interval $(m_{0^{''}}\simeq m_1,\;m_{0^{'}}).$ Therefore, Eq.  
(30) gives
$$C\int _{m_{0^{''}}}^{m_{0^{'}}}dm\;m=9,$$ and hence
\beq
C=\frac{18}{m_{0^{'}}^2-m_{0^{''}}^2}\equiv \frac{18}{\triangle  
}\simeq 27\;
{\rm GeV}^{-2},
\eeq
where the difference $\triangle \equiv m_{0^{'}}^2-m_{0^{''}}^2$ is  
determined
by a distance between the parallel Regge trajectories for the  
$\omega $ and
$\phi $ resonances, which are described by the straight lines  
$J=0.59+0.84M^2$
and $J=0.04+0.84M^2,$ respectively, so that $\triangle \simeq 0.65$  
GeV$^2.$
We note further that the meson nonets may be arranged in the pairs  
of nonets
which have equal parity but different spins (which differ by 2),
e.g.,\footnote{For the scalar meson nonet, we use the $q\bar{q}$  
assignment
suggested by the authors in ref. \cite{linear,enigmas}.}

1 $^3P_0$ $J^{PC}=0^{++},$ $a_0(1320),$ $f_0(1300),$ $f_0(1525),$  
$K_0^\ast
(1430),$

1 $^3P_2$ $J^{PC}=2^{++},$ $a_2(1320),$ $f_2(1270),$ $f_2^{'}(1525),$ 
$K_2^\ast (1430),$ \\

1 $^3D_1$ $J^{PC}=1^{--},$ $\rho (1700),$ $\omega (1600),$ $K^\ast  
(1680),$
(no $\phi $ candidate),

1 $^3D_3$ $J^{PC}=3^{--},$ $\rho _3(1700),$ $\omega _3(1600),$  
$\phi _3(1850),$
$K_3^\ast (1780),$ \\

3 $^1S_0$ $J^{PC}=0^{-+},$ $\pi (1770),$ $\eta (1760),$ $K(1830),$  
(no $\eta ^
{'}$ candidate),

1 $^1D_2$ $J^{PC}=\!2^{-+},$ $\pi _2(1670),$ $K_2(1770),$ (no $\eta  
,\eta {'}$
candidates), \\
and occupy the mass interval of an individual nonet but have 18  
isospin degrees
of freedom in this interval, i.e., twice as much as that for an  
individual
nonet. Moreover, as the temperature gets closer to the critical one  
of chiral
symmetry restoration, we expect the chiral partners (the states  
with equal
isospin but different parity) have equal masses and form parity  
doublets. The
work of DeTar and Kogut \cite{DTK} shows convincingly that the  
``screening
masses'' of chiral partners are different below and become equal  
above a common
$T_c.$ This work was carried out for four quark flavors. Similar  
results were
obtained for two flavors by Gottlieb {\it et al.} \cite{Got}. In these 
calculations, the chiral partners were $(\pi ,\sigma ),$ $(\rho  
,a_1)$ and
$(N(\frac{1}{2}+),N(\frac{1}{2}-)).$ Thus, we expect the correct  
density of
states per unit mass interval to be twice as much as that for an  
individual
nonet, and hence, the correct normalization constant is
\beq
C\simeq 54\;{\rm GeV}^{-2}.
\eeq
In the case we are considering here, one has 6 isospin degrees of  
freedom in
the interval $m_{1/2}^2-m_{0^{''}}^2=\triangle /2,$ in view of (17), 
contributing to the formula for the non-strange condensate, and 3 in the 
interval $m_{0^{'}}^2-m_{1/2}^2=\triangle /2$ contributing to the  
formula for
the strange condensate; i.e., a lower half of a linear spectrum of  
a nonet
contributes to $\langle \bar{q}q\rangle _T,$ while an upper half to  
$\langle
\bar{s}s\rangle _T.$ The corresponding normalization constants are
\beq
C_{\bar{q}q}=\frac{12}{\triangle /2}=\frac{24}{\triangle  
}=\frac{4}{3}C\simeq
70\;{\rm GeV}^{-2},
\eeq
\beq
C_{\bar{s}s}=\frac{6}{\triangle /2}=\frac{12}{\triangle  
}=\frac{2}{3}C\simeq
35\;{\rm GeV}^{-2}.
\eeq
Once the mass spectrum of a nonet (with a given fixed spin) is  
established to
be linear, one may take into account different nonets with  
different spins in
Eqs. (3),(4). As shown in ref. \cite{spectrum}, since the particle  
spin is
related to its mass, $J_i\sim \alpha ^{'}m_i^2,$ $\alpha ^{'}$ being a 
universal Regge slope, the spin degeneracy turns out to be  
proportional to the
mass squared, and the account for different nonets results in the  
following
mass spectrum,
\beq
\tau ^{'}(m)=C^{'}m^3,\;\;\;C^{'}=2\alpha ^{'}C\simeq 90\;{\rm GeV}^{-4},
\eeq
which is the actual resonance spectrum of hadronic matter and leads  
to the
equation of state \cite{EoS}
\beq
p,\rho \sim T^8,\;\;\;p=\rho /7.
\eeq
Bebie {\it et al.} \cite{Bebie} have calculated the ratio $\rho /p$  
directly
from Eqs. (3),(4), with all known hadron resonances with the masses  
up to 2 GeV
taken into account, and found that the curve $\rho /p$ first  
decreases very
quickly and then saturates at the value of $\rho /p\simeq 7,$ as  
read off from
Fig. 1 of ref. \cite{Bebie}, in agreement with (39).

In order to show that the obtained normalization constant is  
correct, we note
that the number of states with the masses up to $M,$ given by the  
mass spectrum
(38), is
\beq
N(M)=\frac{C^{'}M^4}{4}\simeq 22.5\;(M,\;{\rm GeV})^4.
\eeq
For, e.g., $M=1.25$ Eq. (40) gives
\beq
N(1.25)\simeq 55.
\eeq
The masses up to 1.25 GeV have the members of the pseudoscalar and  
vector meson
nonets, and the $h_1(1170),$ $b_1(1235)$ and $a_1(1260)$ mesons,  
the mass of
the latter was indicated by the recent Particle Data Group as 1.23 GeV 
\cite{data1}. We do not include the scalar mesons $a_0(980)$ and  
$f_0(980)$
which seem to be non-$q\bar{q}$ objects \cite{enigmas}, but may  
include the $f_
0(1300)$ meson which has the mass lying in the interval $1-1.5$  
GeV, according
to the recent Particle Data Group. Thus, we have 9+27+1+9+9+1=56  
actual mesonic
species having the masses up to 1.25 GeV, in excellent agreement  
with (41).

For $M=1.7$ GeV, Eq. (40) gives
\beq
N(1.7)\simeq 188.
\eeq
As seen in the Meson Summary Table \cite{data}, the masses up to  
1.7 GeV have
the members of the following nonets: 1 $^1S_0,$ 1 $^3S_1,$ 1 $^1P_1,$ 1 
$^3P_0,$ 1 $^3P_1,$ 1 $^3P_2,$ 2 $^1S_0,$ 2 $^3S_1.$ Therefore, one has 
\beq
(20\;{\rm spin\;states})\times (9\;{\rm isospin\;states})=180\;{\rm  
states},
\eeq
in good agreement with the result (42) given by a cubic spectrum.

For $M=2$ GeV, Eq. (40) gives
\beq
N(2)\simeq 360.
\eeq
The masses up to 2 GeV have the members of all the nonets indicated in 
\cite{data} except for the 1 $^3F_4$ and 2 $^3P_2$ nonets. In this  
case, one
has
\beq
(41\;{\rm spin\;states})\times (9\;{\rm isospin\;states})=369\;{\rm  
states},
\eeq
again in good agreement with the result (44) given by a cubic  
spectrum. Thus,
we consider the cubic spectrum (38) as granted by the actual  
experimental meson
spectrum.

Similarly to the case of an individual nonet, one finds the normalization
constants
\beq
C^{'}_{\bar{q}q}=2\alpha ^{'}C_{\bar{q}q}\simeq 120\;{\rm  
GeV}^{-4},\;\;\;
C^{'}_{\bar{s}s}=2\alpha ^{'}C_{\bar{s}s}\simeq 60\;{\rm GeV}^{-4},
\eeq
and the relations (10),(12) finally take on the forms, respectively, 
\beq
\langle \bar{q}q\rangle _T=\langle \bar{q}q\rangle  
_0\left(1-\frac{T^2}{8f_\pi
^2}-\frac{120\;{\rm GeV}^{-4}\;T^2}{24f_\pi ^2}\int dm\;m^3\frac{m}{T}K_1
\left(\frac{m}{T}\right)
\right),
\eeq
\beq
\langle \bar{s}s\rangle _T=\langle \bar{s}s\rangle  
_0\left(1-\frac{60\;{\rm
GeV}^{-4}\;T^2}{12f_K^2}\int  
dm\;m^3\frac{m}{T}K_1\left(\frac{m}{T}\right)
\right),
\eeq
where we have separated out the contribution of the pions which may  
well be
treated as massless at temperatures $\sim 150$ MeV, and taken into  
account the
remaining hadronic species by the resonance spectrum (38) with the
normalization constants given in (46). We have found that the  
contribution of
the kaons (as well as $\eta $ and $\eta ^{'})$, if separated out as  
well in
Eqs. (47),(48), is comletely negligible at these temperatures. We  
include these
states together with the other mesonic states in the integrals of Eqs. 
(47),(48), which, therefore, have the lower limit of integration  
$\sim 0.5$
GeV. With respect to the remaining integration in (47),(48), we  
note that the
main contribution to integrals of this type is given by the mass  
region in
which $m>>T;$ therefore, one may extend the upper limit of  
integration to
infinity and neglect the lower limit, and obtain, through the  
formula \cite{GR}
$$\int _0^\infty dx\;x^\mu K_\nu (ax)=2^{\mu -1}a^{-\mu -1}\Gamma  
\left(\frac{
1+\mu +\nu }{2}\right)\Gamma \left(\frac{1+\mu -\nu }{2}\right),$$
\beq
\langle \bar{q}q\rangle _T\simeq \langle \bar{q}q\rangle  
_0\left[\;1-\left(
\frac{T}{260\;{\rm MeV}}\right)^2-\left(\frac{T}{220\;{\rm  
MeV}}\right)^6\;
\right],
\eeq
\beq
\langle \bar{s}s\rangle _T\simeq \langle \bar{s}s\rangle  
_0\left[\;1-\left(
\frac{T}{235\;{\rm MeV}}\right)^6\;\right].
\eeq
Temperature dependences of the condensates, as given by (49),(50),  
are shown in
Fig. 1. The critical temperature of the chiral restoration  
transition is $T_c
\simeq 190$ MeV, in agrement with the result of Gerber and  
Leutwyler \cite{GL1}
obtained for the pions and massive states with the nonzero masses  
of $u$- and
$d$-quarks. For the strange quark condensate, the temperature at  
which it is
melted out is $\simeq 235$ MeV, as seen directly in (50). For  
$T\simeq 190$
MeV, Eq. (50) gives $\langle \bar{s}s\rangle _{190}\simeq  
0.72\;\langle \bar{s
}s\rangle _0,$ i.e., only 28\% of the strange quark condensate  
melts by the
chiral restoration transition. It is seen in Eq. (49) that the  
share of the
pions in the overall reduction of the non-strange condensate from  
its initial
value at $T=0$ to zero at $T_c$ is about 53\%, the remaining 47\% is the 
contribution of the massive states. We note also that Eq. (49) may  
be well
approximated by
\beq
\langle \bar{q}q\rangle _T\simeq \langle \bar{q}q\rangle  
_0\left[\;1-\left(
\frac{T}{190\;{\rm MeV}}\right)^3\;\right].
\eeq
The formulas (49),(50) have been obtained for the meson resonances  
alone. There
is no difficulty of principle to consider the baryon resonances in  
a similar
way, with the inclusion of two chemical potentials, for both  
conserved net
baryon number and strangeness. As we have checked in ref.  
\cite{spectrum}, a
mass spectrum of the $SU(3)$ baryon multiplets is linear, as well  
as for the
meson nonets, although to establish its correspondence to the  
Gell-Mann--Okubo
formulas is more difficult than for a meson nonet, since these  
formulas are
linear in mass for baryons (more detailed discussion is given in ref. 
\cite{spectrum}). Recent result of Kutasov and Seiberg \cite{KS}  
shows that
the numbers of bosonic and fermionic states in a non-supersymmetric 
tachyon-free string theory must approach each other as increasingly  
massive
states are included. The experimental hadronic mass spectrum shows  
that in the
mass range $\sim 1.2-1.7$ GeV, the number of baryon states nearly  
keeps pace
with that of meson states \cite{FR} (and, therefore, is well  
described by the
same cubic spectrum as for the mesons, Eq. (38)). Above $\sim 1.7$  
GeV, the
number of the observed baryons begins to outstrip that of the  
mesons, and then
greatly surpasses the latter at higher energies (indicating,  
therefore, that
the baryon resonance spectrum grows faster than (38) in this mass  
region, since
the cubic spectrum (38) describes the meson resonances well, up to,  
at least,
2 GeV, as we have seen in Eqs. (41)-(45)). The explanation of this  
behavior of
the experimental resonance spectrum was found by Cudell and Dienes  
in a naive
hadron-scale string picture \cite{CD}: the ratio of the numbers of  
the baryon
and meson states should, in fact, oscillate around unity, with the mesons
favored first, then baryons, then mesons again, etc. Keeping in mind this
picture, we may assume that the ``in-average'' baryon resonance  
spectrum has
the same form, Eq. (38), as the meson resonance one. It is then  
possible to
estimate the contribution of the baryon resonances to the temperature 
dependences of the quark condensates by assuming that the formulas  
(26),(27)
hold for the baryon resonances, as well as for meson ones, and that  
out of 27
degrees of freedom of $SU(3)$ baryon octet, nonet and decuplet,  
2/3, i.e., 18,
contribute to $\langle \bar{q}q\rangle ,$ and 1/3, i.e., 9, to  
$\langle \bar{s}
s\rangle .$ If one now neglects, for simplicity, the baryon number and 
strangeness chemical potentials (i.e., considers the case of both  
zero net
baryon number and strangeness), and takes into account the baryon  
resonances
along with the meson ones by the mass spectrum (38), one will  
obtain the same
formulas, (47),(48), but with the factors in front of integrals  
which are twice
as much as those in the case of the meson resonances alone. These  
formulas will
further reduce to the relations
\beq
\langle \bar{q}q\rangle _T\simeq \langle \bar{q}q\rangle  
_0\left[\;1-\left(
\frac{T}{260\;{\rm MeV}}\right)^2-\left(\frac{T}{195\;{\rm  
MeV}}\right)^6\;
\right],
\eeq
\beq
\langle \bar{s}s\rangle _T\simeq \langle \bar{s}s\rangle  
_0\left[\;1-\left(
\frac{T}{210\;{\rm MeV}}\right)^6\;\right],
\eeq
shown in Fig. 2. One sees that now the critical temperature of the  
chiral
symmetry restoration transition is $T_c\simeq 175$ MeV, while the  
strange quark
condensate melts out at $\simeq 210$ MeV, and about 34\% of the  
latter melts by
the chiral restoration transition. Now the pions' contribution to  
the overall
reduction of the value of the non-strange condensate is about 45\%, the 
remaining 55\% is the share of the massive states.

\section*{Concluding remarks}
We have derived the formulas for the temperature dependences of the  
quark
condensates by taking into account the contribution of the massive  
states
parametrized by a resonance spectrum which has been established by  
the authors
in previous papers. In the case of the meson resonances alone, our  
results
agree those obtained previously by Gerber and Leutwyler, and  
suggest the chiral
symmetry restoration transition to occur at the critical temperature 
$T_c\simeq 190$ MeV. With the baryon resonances taking into account  
along with
the meson ones, the critical temperature turns out to be $T_c\simeq  
175$ MeV,
which is closer to the currently adopted value $140-150$ MeV  
established by
lattice gauge calculations \cite{Ber}. In either case, only $\sim  
30$\% of the
strange quark condensate melts by the chiral restoration transition, in
agreement with lattice calculations \cite{KSW}. The share of the  
pions in the
overall reduction of the non-strange condensate from its initial  
value at $T=0$
to zero at $T_c$ is $\sim 50$\% in either case, the remaining $\sim  
50$\% is
the contribution of the massive states.

\section*{Acknowledgements}
One of us (L.B.) wish to thank E.V. Shuryak for very valuable  
discussions on
hadronic resonance spectrum.

\newpage
\centerline{FIGURE CAPTIONS}
\bigskip
\bigskip
\bigskip
\bigskip
\hfil\break
Fig. 1. The ratio $\langle \bar{q}q\rangle _T/\langle  
\bar{q}q\rangle _0$ as a
function of temperature for a) non-strange, b) strange quark  
condensate, in
the case of the meson resonances alone, as given in Eqs.  
(49),(50).\hfil\break
\hfil\break
\hfil\break
\hfil\break
Fig. 2. The same as Fig. 1 but in the case of both the meson and baryon
resonances, as given in Eqs. (52),(53).
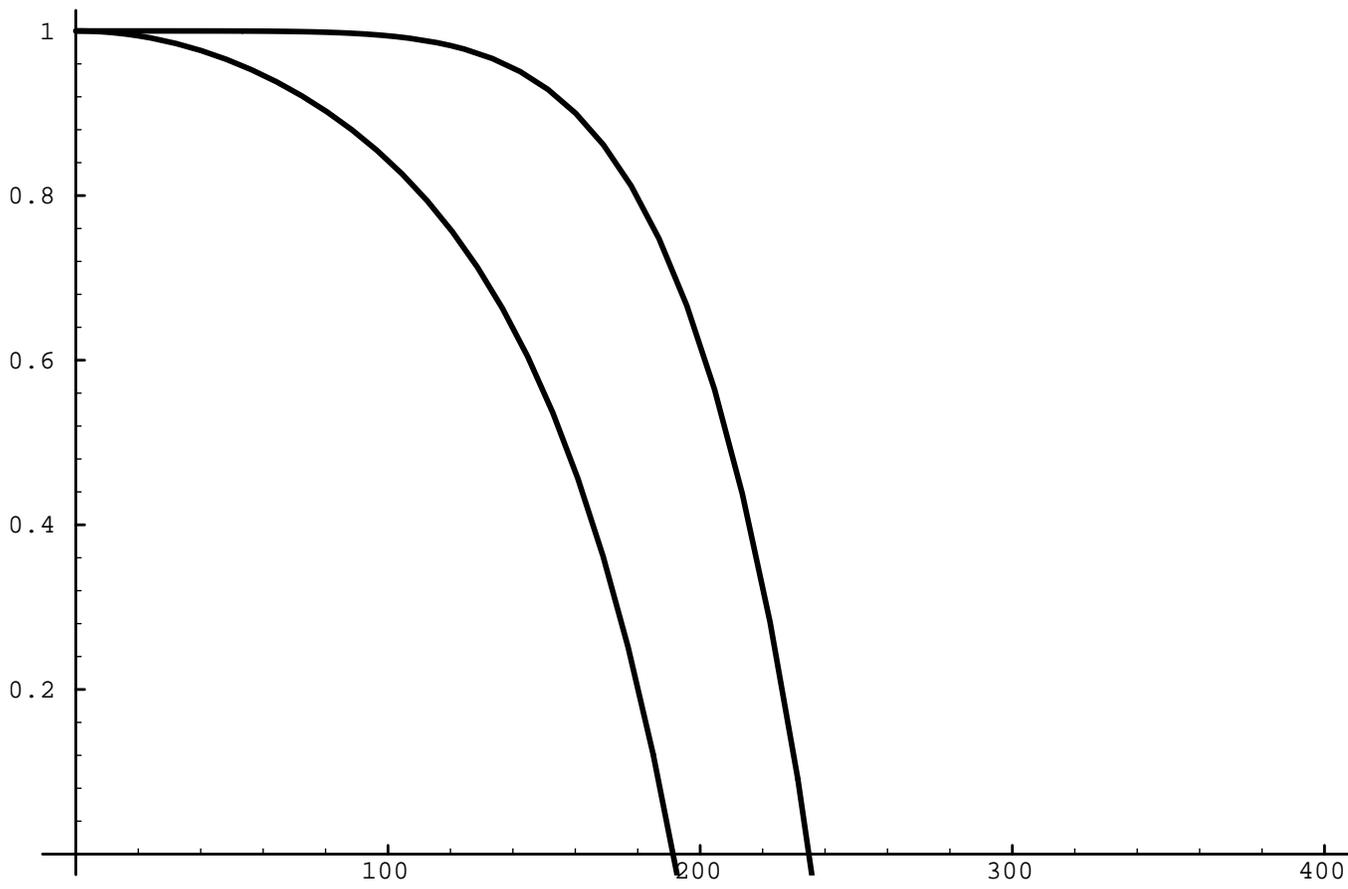
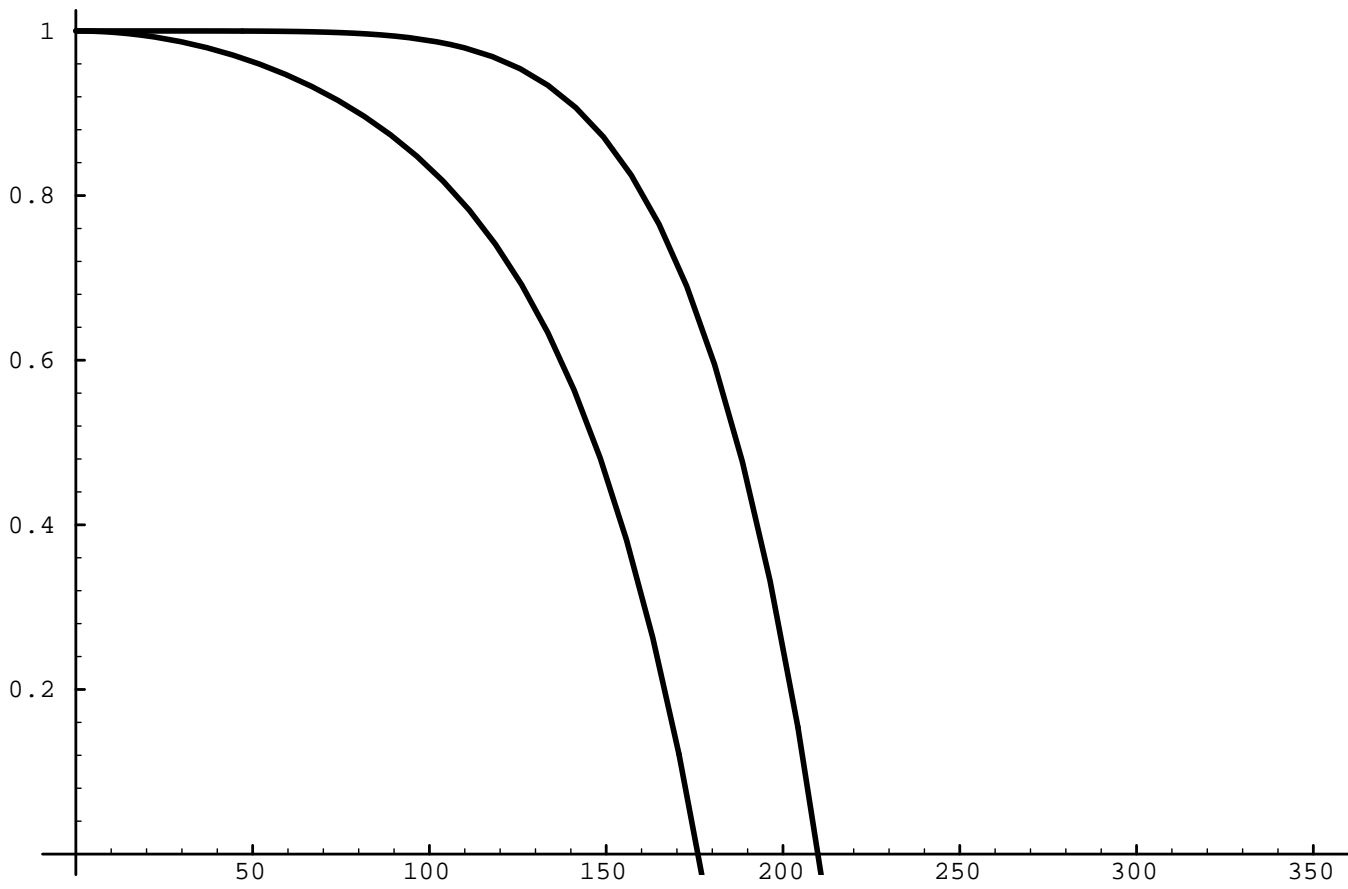
\end{document}